# User Behavior Discovery in the COVID-19 Era through the Sentiment Analysis of User Tweet Texts

Amin Mahmoudi[1], Victoria Wai-lan Yeung[2,3], Eric W. K. See-To[4]

[1] Department of computing and decision sciences, Lingnan University, Hong Kong, aminmahmoudi@ln.edu.hk
[2] Department of Applied Psychology, Lingnan University, Hong Kong, vickiyeung@ln.edu.hk
[3] Wofoo Joseph Lee Consulting and Counselling Psychology Research Centre, Lingnan University, Hong Kong
[4] Department of computing and decision sciences, Lingnan University, Hong Kong, ericseeto@ln.edu.hk

**ABSTRACT**

The coronavirus disease (COVID-19) outbreak was declared a pandemic in March 2020 and since then it has had a significant effect on all aspects of life. Although we live in an information era, we do not have accurate information about this disease. Online social networks (OSNs) play a vital role in society, especially people who do not have trust in the government would tend to have more confidence in the evidence that is formed by social networks. The advantages of OSNs in the COVID-19 era are clear. For instance, social media enables people to connect with each other without the need for real-world face-to-face social interaction. Social media networks also act as a collective intelligence in the absence of world leadership. Therefore, in this study, considering the phenomenon of information diffusion in OSNs, we focus on the effects of COVID-19 on user sentiment and show the user behavior trend during the early months of the pandemic through mining and analyzing OSN data. Moreover, we propose a data-driven model to demonstrate how user sentiment changes over a period of time and how OSNs help us to obtain information on user behavior that is very important for the accurate prediction of future behavior. For this purpose, this study uses tweet texts about COVID-19 and the related network structure to extract significant features, and then presents a model attempting to provide a more comprehensive real picture of current and future conditions.

## 1. INTRODUCTION

*A. Motivation*

The world today is facing a huge yet invisible enemy, a coronavirus known as COVID-19, and it seems that governments do not have any specific strategy to combat and overcome this disease. It seems that the main reasons that governments have been unable to respond quickly to this crisis are (1) lack of global leadership, (2) lack of knowledge about the disease, and (3) lack of knowledge about people's behaviors toward and practices against the disease. In this study, we discussed the possibility of predicting human behaviors in the current and post-COVID era by identifying social reactions and the prominent features which derived the society during the COVID-19 pandemic.

Online social networks (OSNs), such as Twitter and Facebook, are one of the best platforms for identifying social behaviors in this challenging time. Undoubtedly, the main difference between this pandemic and the previous one over one hundred years ago is that the current pandemic coincides with a time when society widely uses OSNs. Although some disadvantages of using OSNs have been observed, such as the spreading of rumors and fake news,

OSNs play a critical role in our society, especially when governments do not provide accurate statistics of the toll that COVID-19 is having on human life. When countries do not deliver accurate statistics related to COVID-19 for whatever reason, this could lead to misunderstandings regarding the speed or any actions taken to fight the disease. Besides, the information propagated on OSNs is the leading source of evidence for people who do not have trust in the government. The enablement of social interactions that form a network is another significant benefit of OSNs. The information on social interaction as digital content (e.g., film, text) is very beneficial in data mining techniques, and the network structure helps us to understand the social relationship. According to the social physics theory (Pentland, 2015), it is possible to determine social behavior from the analysis of big data (which formed in OSNs) and thereby formulate social events. A social network is a rich resource from which to extract social behavior and it is possible to find patterns hidden in the huge amount of data generated in such a complicated network (Bhattacharya & Kaski, 2018). Another target of social physics is the finding of social norms. Therefore, the analysis of OSNs contents may facilitate faster and more accurate actions to fight the current pandemic in comparison with the previous pandemics. Indeed, it has been argued that OSN analysis is one of the main tools that can be used to identify social sentiment and social capital. Social sentiment refers to emotions behind society and social capital "comprises social networks, norms of reciprocity or social support and social trust" (Ferlander, 2007). Both social sentiment and social capital are especially important at this difficult time when a crisis happened (Pitas & Ehmer, 2020). To elaborate, if we can know people's emotions and social interactions via analyzing their communication contents on OSNs, we can have a better and more accurate prediction of their behaviors because some behaviors are emotionally-driven (Pêcher et al., 2011). So, by understanding people's emotions, we can understand or predict their behaviors, and this can subsequently help governments to develop appropriate strategies in response to people's emotions and behaviors during the pandemic.

**B. Contributions**

This study aims to analyze people's social interactions on an OSN and the network structure of that OSN in order to reveal people's behaviors during the occurrence of a pandemic. To this end, we used Twitter text data gathered during the COVID-19 era and conducted a sentiment analysis on these data to provide a more comprehensive picture of society during this time. Moreover, we analyzed the role of social capital, especially focusing on influential and verified users around the world, by extracting related features using a correlation matrix. We also highlighted how socialization can help to inject hope into society and raise morale in times of crisis. Finally, we searched tweet texts to identify some important themes for a better understanding of social change for use in recommender systems in a pandemic era. The main contributions of this study are as follows:

1. It reveals how society behaves during a pandemic, and shows that, contrary to the popular belief that people in a given society follow the most-followed people (influencers), there is no relationship between user sentiment and influencers (as a network structure) during a critical situation such as a pandemic. In other words, user sentiment is more a function of time rather than network structure.
2. It shows the relative importance of key societal issues at different time intervals during a pandemic. By identifying and analyzing the data on five main themes in the COVID-19 era, this study reveals which theme is the most important for people in a given society. These results can be used in marketing, recommender systems, and urban planning.

The rest of this paper is organized as follows: In the next section, the related work on this domain is reviewed. Then, in section 3, the preliminaries and notations are provided. This is followed by section 4 in which the methodology is described in detail. In section 5, the experimental results are presented. Finally, in section 6, conclusions are drawn.

## 2. RELATED WORKS

The world has faced many pandemics throughout history from the Antonine Plague to COVID-19. Although humans have experienced several pandemic diseases over the past centuries, there is a difference between the current one and those previous ones in that the current COVID-19 pandemic is happening in the OSN era.

During the COVID-19 pandemic, some data scientists have applied sentiment analysis to texts propagated via OSNs (Dubey, 2020; Barkur et al., 2020). Sentiment analysis is a subfield of natural language processing, which in turn is a sub-domain of artificial intelligence. Sentiment analysis is used to extract the polarity of each sentence in communication and categorize it into three subgroups, i.e., positive, negative or neutral (Kharde & Sonawane, 2016). Nowadays, the ability to recognize the sentiments and opinions of users' in, for example, review texts on OSNs plays a critical role in the effectiveness of marketing, recommender systems, and the organization of social events (Feldman & Ronen, 2013; da Silva et al., 2016). Recently, many researchers have introduced new techniques and methods to improve the accuracy of user sentiment identification regarding a specific topic in OSNs. These techniques and methods are used to analyze the sentiments expressed in comments at three levels, namely, the document, sentence, and word level. They include the support vector machine, naïve Bayes, and association rules. Besides, corpus-, lexicon- (Taboada et al., 2011) and dictionary-based methods can also be used in word sentiment classification (Giachanou & Crestani, 2016). Yoo et al. (2018) used deep learning techniques for sentiment analysis. However, the majority of the methods in the literatures have primarily focused on the content without considering the characteristics of the users who post the content.

Jin and Zafarani (2018) showed that it is possible to predict user sentiment by analyzing the relationships within a social network. In their work, they used structural properties at four levels, namely, the ego, triad, community, and network level, to predict sentiments. This approach enabled them to analyze the effects of friend sentiments on user sentiments. Tan et al. (2013) applied the concept of the social relationship to improve user-level sentiment analysis. The key assumption behind their idea is that users with a mutual friendship are more likely to have similar opinions. Their study involved Twitter data and they focused on the user level rather than the tweet level because they considered that the tweet level is error-prone and time-consuming. To solve the problem of language variation in sentiment analysis, Yang and Eisenstein (2017) presented a method based on the notion of social attention. In a similar vein, Yang et al. (2016) used a sociological theory, namely, homophily, to overcome the ambiguity problem in Twitter text, which is an entity link system. The theory of homophily asserts that users who are connected are more likely to share similar behaviors and common interests.

Unlike most existing studies that relied on text information to do sentiment analysis, Tang et al. (2015) used user and product information to do sentiment analysis. They introduced a model, called the user product neural network, to extract user and product information for use in the sentiment classification of documents. Gui et al. (2017) also leveraged user or product information as a feature with text information to train a sentiment classifier. They considered four types of relation to make heterogeneous networks, namely, the word relation, word polarity relation, word user relation, and word product relation.

On the other hand, Gong and Wang (2018) applied self-consistency theory to analyze user behavior in social media. They assumed that different users may share the same intents. Their proposed method consists of two main models: a logistic regression model to map the textual contents to the sentiment polarity, and a stochastic block model to capture the relatedness among users. Zou et al. (2018) designed a new method based on social and topic content to identify user sentiment. Fornacciari et al. (2015) combined the analysis of the social network structure with sentiment analysis to reveal the incorrect results of sentiment analysis based on the network topology. In addition, they demonstrated that polarity of feeling can lead to the identification of the semantic connections in a network. Additionally, West et al. (2014) used linguistic and social features to predict people's opinion of one another.

Recently, several studies have emerged investigating the role and value of OSNs in the COVID-19 era. For instance, Pitas and Ehmer (2020) discussed the impact of social capital on crisis management process, and recommended that governments, communities, and individuals should maintain social capital in the COVID-19 era. They also recommended that social media sites as contemporary platforms should generate social capital. Along similar lines, Chuang et al. (2015) evaluated the influence of social capital on the response to the influenza pandemic. In the current context, Limaye et al. (2020) focused on the guiding role of social media as a new digital reality, while, more specifically Kim (2020) conducted a semantic social network analysis to reveal the effect of social grooming on incivility in the COVID-19 era, the results of which showed that network size is related to content-expressive behaviour. Moreover, Wiederhold (2020) focused on the issue of mental health care during the COVID-19 pandemic and reported that social networks relieve stress and provide an opportunity to share feelings. On the other hand, Cinelli et al. (2020) addressed the broader issue of the diffusion of information about COVID-19 on OSNs and provided an assessment of each source of information. In regard to the influence of social media, Depoux et al. (2020) mentioned its harmful effects. Finally, in a similar vein, Rosenberg et al. (2020) discussed the role of Twitter during this challenging time and categorized tweets into two types, as either harms or benefits. Dubey (2020) conducted a sentiment analysis of tweets by users in 12 countries and discussed the emotions associated with such tweets. Their results clearly showed the crucial role of social networks in both the offline and online states in times of upheaval, and how sometimes the effect is negative and harmful and sometimes useful.

3. **PRELIMINARIES AND NOTATIONS**

This section provides the definitions of the three main concepts that are the basis of the current study, they are OSN, sentiment and OSN structure.

*Definition 1 (Online Social Network).* An OSN is an online service that facilitates communications in a social network. Each OSN can be represented as $G(U, C, T)$, where U is a set of user IDs, $C$ is a set of triplets of the form $(u_i, u_j, t)$, where $u_i, u_j \epsilon U$, $t\epsilon T$, and $T$ are a set of timestamps in which two users $u_i$ and $u_j$ communicate with each other. According to the above definition, an OSN is a time-varying and dynamic domain in which the behavior of the network changes over time (Mahmoudi et al., 2020).

*Definition 2 (Sentiment).* The Cambridge dictionary defines sentiment as "a thought, opinion, or idea based on a feeling about a situation, or a way of thinking about something" (Dictionary, 2020). Sentiment analysis studies try to identify emotions of humans from review texts input via a computer or other Internet-enabled device. In this study, we measured the emotions of users in OSNs by computing the sentiment score of each tweet text for any specific topic. For each user who tweeted on a topic, a sentiment score vector was calculated and represented as $S_I = \{s_1, s_2, \ldots, s_n\}$, where $s_i$ represents the $i-th$ sentiment score of user $I$ who tweeted on a specific topic. Each user has a number of tweets in the network. $TT_i = \{TT_1, TT_2, \ldots, TT_j\}$ represents a tweet on a topic by user $I$, as formulated in Equation 1:

$$S_I = Sentiment\ (TT_j)\ , if\ TT_j\ is\ tweeted\ by\ user\ I \qquad (1)$$

Here, the sentiment is a function that computes the sentiment score of a tweet text, and the sentiment score can be a positive or negative value.

*Definition 3 (OSN Structure).* In this study, OSN structure refers to network degree, where the degree can be described as the number of friends in an OSN (Nettleton, 2014). This metric ranks the nodes based on their connections, as shown in Equation 2 (Freeman, 1978):

$$C_d(u_i) = d_i, \qquad (2)$$

where $d_i$ is the degree of $u_i$, where $u_i$ is user $I$ in an OSN

*Definition 4 (user behavior).* The user behavior is keyword in this study. It refers to how users react to various themes on Twitter and related trends that users follow on this social network.

In this study, we considered the number of friends and the number of followers as the network structure. In addition, we viewed network properties as related to the temporal and dynamic nature of the network. Thus, this study considered the lifespan of the network as a network property.

## 4. METHOD

This study has deployed the social physics theory (Pentland, 2015) to formulate the relationship between the significant characteristics of network information and users' sentiments to reveal a more comprehensive picture of the COVID-19 era and predict user sentiments after the COVID-19 pandemic is over. As it is particularly important to try to identify social norms during this challenging time, in this work, we tried to apply the ideas of social physics theory to real-world data in order to find a way to predict user behavior and find similar patterns in each time interval of an OSN dataset. Hence, we analyzed around 18 million tweet texts related to COVID-19. We deployed a data-driven approach to reveal the current condition and to predict user behaviors beyond the COVID-19 era. This study consisted of two main phases: (1) drawing out the relationship between network information and people's sentiment, which represents the sociological and psychological behavior of users (human beings) and (2) analyzing the user behavior in tweet texts to develop a data-driven method with the aim of predicting user behaviors in the future.

**4.1 Data**

The tweet dataset that was used in this study contained the tweets of users who applied the following hashtags: #coronavirus, #coronavirusoutbreak, #coronavirusPandemic, #covid19, and #covid_19. The dataset also included tweets that incorporated the following additional hashtags: #epitwitter, #ihavecorona" (Kaggle, 2020). The tweets were gathered for the period from March 4 to April 15, 2020. The collected dataset contained approximately 18 million tweets, as described in Table 1.

Table 1. Dataset statistics

| Dataset | No. of Users | No. of Tweets | English Tweets |
|---|---|---|---|
| **Coronavirus (covid19) Tweets** | 4,373,783 | 17,900,881 | 9,457,721 |

The main important fields of this dataset are user ID, text, date of creation (of tweet), favorites count, retweet count, followers count, friends count, account created at (date), verified (user), and language. We selected English-language tweet texts only. The "account created at" field was used to compute the user lifespan by calculating the number of days between creating an account and sending a tweet related to COVID-19.

**4.2 Feature Extraction**

We extracted the important features that affected people's (users') sentiments and emotions during the COVID-19 pandemic. The first step involved identifying the characteristics that affect user sentiments. To do this, we created a correlation matrix of the significant factors including (1) the number of friends (friends count), which reflects the sociality of the user, (2) the followers count, which indicates the extent to which the user is able to influence other users, (3) the favorites count, which provides the number of times the tweet has been favorited, (4) the user lifespan, which shows how long the user is a member of the network, and (5) verified status, which reflects authentic, notable, and active users in Tweeter (see Table 2).

Table 2. Symbols of main user characteristics

| User Characteristic | Symbol |
|---|---|
| Favorites count | $fa$ |
| Friends count | $fr$ |

| Follower count | *fo* |
| --- | --- |
| verified | *v* |
| lifetime | *li* |
| User behaviour | *UB* |

To compute the sentiment score, we used the "*sentiment_by*" function of the 'sentimentr' package of R programming language, where the output of this function is a data frame that consists of "avg_sentiment" that shows the sentiment (polarity) mean score. *Sentiment_by* calculates the polarity (sentiment) of text by grouping words (Rinker, 2019). Two main arguments of this function are:

*Text.var.* It is the first argument, which refers to the text we tried to find its polarity. We used the user Tweet text for this argument.

*averaging.function.* It refers to a method that the *sentiment_by* used to perform group average. This argument can take three different methods as *average_downweighted_zero*, *average_weighted_mixed_sentiment*, and *average_mean*. Due to this study has been conducted to identify positive and negative sentiment in the COVID-19 era, we tried to down weight neutral sentiment (zero value). Therefore we used *average_downweighted_zero* as averaging function.

*Sentiment_by* returns a data frame consisting of *word_count* that indicates a count of word grouping by considering particular words in the text, *sd* indicates the standard deviation of polarity (sentiment) value, and *ave_sentiment* that indicates the mean value of polarity in each group. We used *avg_sentiment* to assign sentiment value to each tweet text since we wanted to show the average value of sentiment.

Figure 1 shows the correlation matrix for the sentiment score and user characteristics (lifespan, favorites count, followers count, friends count and verified status). There are two main types of correlation coefficients, namely, the Spearman and the Pearson coefficients (Schober et al., 2018). In this phase, we used Spearman correlation because it is suitable for ordinal data; the tweets dataset contained verified and sentiment (emotion) variables, which belong to this type of data. Therefore, the Spearman's rank correlation was used to measure the association between the aforementioned user characteristics and the sentiment score. A correlation coefficient value can fall between −1 and +1, where a greater association with values of +1 or −1 indicates a perfect association. A correlation coefficient value close to +1 indicates positive relationship between two variables and a value close to -1 shows that there is a negative relationship between the two variables. A correlation coefficient of zero would indicate that there is no association between the two variables.

Figure 1 shows that the correlation between sentiment and lifespan is around -0.01; this value is similar to the correlation coefficient between sentiment and followers count. Although the correlation coefficient of friends count is relatively higher, there is a negligible association as the value is -0.05. On the other hand, the correlation coefficient between sentiment and verified user has a negligible positive value of +0.04. Generally, Figure 1 shows that in times of crisis, not only are people not affected by influential users, there is no difference between ordinary users and significant users (influencers, verified users). Hence, it can be observed that there is no robust relationship between user sentiment (the score denotes the sentiment value) and user characteristics.

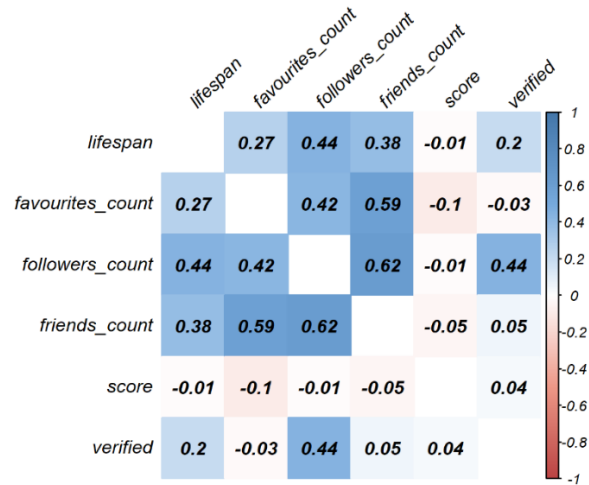

Fig. 1. Correlation matrix between user sentiments and characteristics

Table 3 provides a narrow (microscopic) picture of the relationship between some of the main user characteristics and sentiment scores. It should be noted that the mean value of the sentiment score is 0.017. It can be observed that the sentiment score for verified users is 0.029, which is higher than the mean value of the sentiment score for all users. This means that, generally speaking, in critical situations verified users posted more positive tweets and were more relaxed than the rest of society. In contrast, the tweets posted by the most-followed users and those with high degree centrality (which was computed based on the number of friends) were more negative than the average of the entire statistical population by the value of 0.001 and 0.0064, respectively. At the same time, the sentiment value of users whose activity time was longer than average level was close to the average sentiment value of the entire statistical population, where the sentiment value for this group was 0.0165 and the mean value of the lifespan (duration of activity) was 2478 days. However, there was a reverse relationship between users with the highest favorited tweets and their sentiments, meaning that as the number of likes of a certain tweets increased, their sentiment value decreased. The sentiment score for this group of users was -0.003, where the mean value for this property was 13260.

Table 3. Relationship between mean value of user characteristics and sentiment score

| Characteristics | $v$ | $fo$ | $fr$ | $li$ | $fa$ |
|---|---|---|---|---|---|
| mean value | 82528 (total verified users) | 53584 | 2184 | 2478 | 13260 |
| Sentiment score | 0.0294 | 0.001 | 0.0064 | 0.0165 | -0.003 |

As shown in Table 3, user sentiment is merely a function of time, where the network characteristics that are shown in Fig. 1 do not have an effect on society (macroscopic) sentiment in a critical situation such as the COVID-19 pandemic. So it can be said that

$$S_{critical\ situation} = f(t), \tag{3}$$

where $S$ is the sentiment score and a function of $t$, where $t$ denotes time.

**4.3 User Sentiment Analysis**

Figure 2 shows the user sentiment in the period under studied (i.e., from 04-March-2020 to 15-April-2020). As can be seen, in the early days of the pandemic until 13[th] March 2020, the sentiment values are negative, which shows that people were very concerned about COVID-19. Then, from March 13[th] onwards, the sentiment values become more positive, which implies that people were coping with the stressful situation.

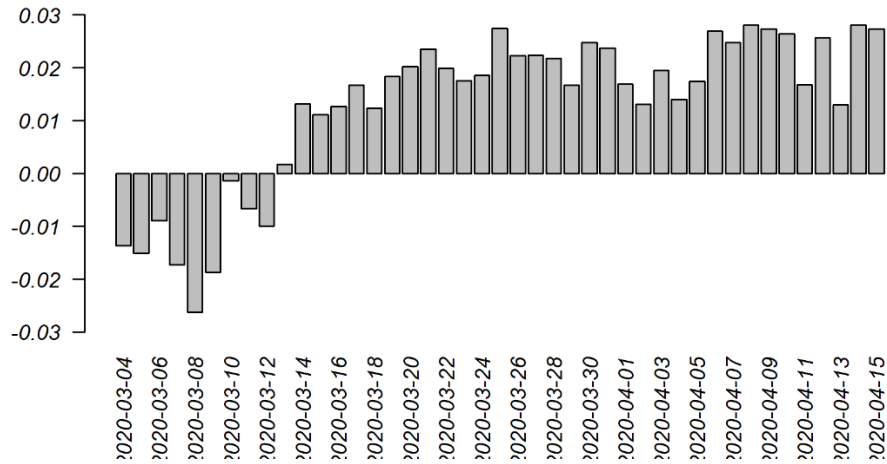

Fig. 2. Sentiment values of tweet texts during the time period from 04-March-2020 to 15-April-2020

**4.4 User Behavior Analysis**

In this section, we analyzed user tweet texts about the various challenging themes that arose during the studied period to determine whether there were any changes in user behaviors during the pandemic. For this analysis, we selected five themes widely mentioned during the pandemic, namely, shopping, personal hygiene, entertainment, protest, and employment (job) (Fegert et al., 2020)(see Table 4). These five themes were chosen because they are considered to represent the main challenges and concerns of people all around the world during the pandemic. First of all, shopping is one of the main concerns. It is important for retailers, manufacturers and the wider economy to understand user behaviors, especially consumption behaviors and purchasing intentions during the pandemic. Second, personal hygiene is another crucial theme. Public health officials in different countries around the world are very concerned about ensuring that their populations maintain and improve good personal hygiene during the COVID-19 pandemic because it plays a key role in reducing the spread of disease (WHO, 2020). The third key theme is entertainment. Quarantine, a strategy that has been implemented in many countries during the pandemic, can lead to depression (Fegert et al., 2020), and entertainment is seen as factor that can alleviate depression and isolation. The forth main theme is protest. Protest occurred in many societies, especially those that are subjected to pressures such as pandemics and other associated social factors. Hence, identifying user behaviors and trends regarding this theme is crucial for governments' governance and for society stability. The fifth and last theme is job. Many people have lost and will continue to lose their jobs due to COVID-19. It is important to examine how people behave and cope with the consequences related to this important theme. In short, the extent and type of user behaviors in social networks with regard to these five themes would signal the level of priority given to these themes by users. Any changes in user behaviors would indicate how people face the challenges. For each theme, we identified some close synonyms or keywords (see Table 4).

Table 4. Keywords for the five themes

| Themes | Keywords |
| --- | --- |
| Shopping | 'shop'- 'shopping'- 'buy' - 'buying' - 'market' - 'store' - 'purchase' |
| Personal hygiene | 'disinfectants' - 'hand' - 'wash' - 'bactericide' - 'cleanser' - 'antiseptic'- 'cleaning agent' -'clean' - 'detergent' - 'soap' - 'suds' - 'lather' |
| entertainment | 'party' - 'celebration' - 'cheer' - 'game' - 'sport' - 'relaxation' - 'entertainment' - 'relief' -'fun' - 'enjoyment' - 'pastime' - 'picnic' |
| protest | 'protest' - 'demonstration' - 'dissent' - 'objection' - 'outcry' - 'revolt' - 'manifestation' - 'turmoil' - 'violence' |
| job | 'job' - 'position' - 'profession' - 'career' - 'vocation' - 'appointment' - 'work' |

Figure 3 shows the word frequencies per day for the five aforementioned themes during the period 04-March-2020 and 15-April-2020. Each theme consists of six plots that depict the word frequency of all users, the word frequency of users who have a favorites count, friends count, followers count and lifespan value higher than the mean value, and the word frequency of users who are verified tweeters.

Before analyzing the frequency of each theme, it should be emphasized that the low frequency observed is due to the number of words this study considered for each theme. On the other hand, although we removed a vast variety of non-essential words such as prepositions, pronouns, etc., some others remained in the users' tweet texts, which affected the frequency of the words for each theme. Nevertheless, it is considered that the frequency ratio still provides a reasonably accurate picture of reality.

*Shopping*. Figure 3 shows that at the beginning of the pandemic when people have not much information about the diseases and when panic bulk buying occurred, there was a sharp increase in word frequency search related to the theme of shopping (word frequency was 0.002). However, as time passed, the shopping theme among all users shows a descending trend. The word frequency value fell to less than 0.001 in the middle of the studied period, and to around 0.0009 at the end of the studied period. This sharp decrease pattern indicates an over 50% fall (percentage decrease) in word frequency related to the theme of shopping. For the other user characteristics, this trend is also, overall, a descending one, although there are not completely same.

*Personal hygiene.* Figure 3 shows that all users are very concerned about hygiene issues as compared to any other themes with a maximum frequency value of around 0.0029, although this issue loses its significance over time. The frequency value of the personal hygiene theme for the first day of tweeter pandemic data (4 March, 2020) is 0.0028, while the value for the last day of tweeter pandemic data (14 April, 2020 is around 0.0006, which shows that there is a more than 78% fall in usage. This trend is similar among other types of user characteristic, i.e., 0.0005 and 0.00008 for users with a higher favorites count than the mean value, 0.0001 and 0.00003 for most-followed users, 0.0004 and 0.0001 for users with a high degree centrality value, 0.0015 and 0 for experienced users, and 0.0001 and 0.00007 for verified users. This data indicates that at the beginning of the pandemic when people do not have much information about the disease, they relied on the social networks to gather information especially those information about personal hygiene. As time passed, when people gained more knowledge concerning the disease, they became more relaxed which is reflected by the lower word frequency associated with the theme of personal hygiene.

*Entertainment.* In the first days of the pandemic, people are less concerned about entertainment but more concern about how to tackle the pandemic, as such we can observe that the value of the word frequency for the theme of entertainment is just around 0.0005 at the beginning of the studied period. Nevertheless, as stated in the literature, the longer people need to struggle with and suffer from the associated consequences of the pandemic, such as lockdown, quarantine, keep social distance and being socially and physically isolated, the more likely they would develop loneliness or depression(Fegert et al., 2020), and the more likely they need to find entertainment to cope with or distracted from these problems. We can observe that along the timeline, people try to cope by sharing more entertainment theme in their social networks (word frequency value generally higher than 0.001). This is particularly true among the longer lifespan users – the longer lifespan users showed a peak in word frequency about entertainment in the middle of the studied period compared to other regular users. The data for the last day for the lifetime measure are not available. The plots of the other four types of user characteristic are perfectly compatible. Overall, people show a steady high word frequency pattern of the entertainment theme as long as the pandemic is still bothering them.

*Protest.* The word frequency pattern of protest theme is quite the opposite of that of the shopping and personal hygiene themes. On the first day, the word frequency for protest is 0.00008 but it later sharply increased to 0.0002 on the last day. This shows a more than 150% increase for the protest theme during the studied period. Perhaps in addition to quarantine policy it was due to that people started to realize that the government is incapable to tackle the disease and felt hopeless, leading to more protesting activities in the societies (Carothers, 2020) and subsequently sharing more words related to the theme of protests. All users' types shared the same pattern until the middle days of the studied period. The word frequency value on the first day for the most-followed users is from 0.0 at the beginning, to 0.00002 in the middle (20 March, 2020) and falls to 0.000016 at the end of the studied period). For users with high degree centrality (friends count), the respective values are 0.000012, 0.00006, and 0.000054. The values are 0.00005, 0.000062, and 0.00005 for users with a high favorites' tweets and 0.00006, 0.0001, and 0.00008

for experienced users. Despite the fact that the verified users' word frequency is similar to that of all other users, the values for the verified users are 0.000012, 0.000011 and 0.00004. Based on these results, it is expected that the word frequency of the protest theme will continue to increase in the future according to the longevity of the crisis, which may have implications for society stability.

*Job.* The word job is a keyword that has high frequency throughout the studied period. A few days after the existence of a pandemic is confirmed, it becomes a hot keyword and the main issue discussed among tweeters. The maximum frequency level of this theme is more than 0.0026, and it remains an important concern for a long time as compared with the personal hygiene theme that is the main concern in the early days of the pandemic. On the first day of the pandemic, the frequency value for job is 0.0014 and at the end of the experiment period it is around 0.0022. Also, it can be seen that there is no consistency between the word frequency of all regular users and the word frequencies of other users (five users types), especially those who are the most-followed, those who have the highest favorites count, those who are verified users and those who have high degree centrality. Over the period under studied, the frequency value of the job theme remains steady and is the main issue discussed in the tweets. Due to the critical situation and the bankruptcy notices issued by some giant companies, it is clear that people are concerned about their jobs and the consequent difficulties they could face.

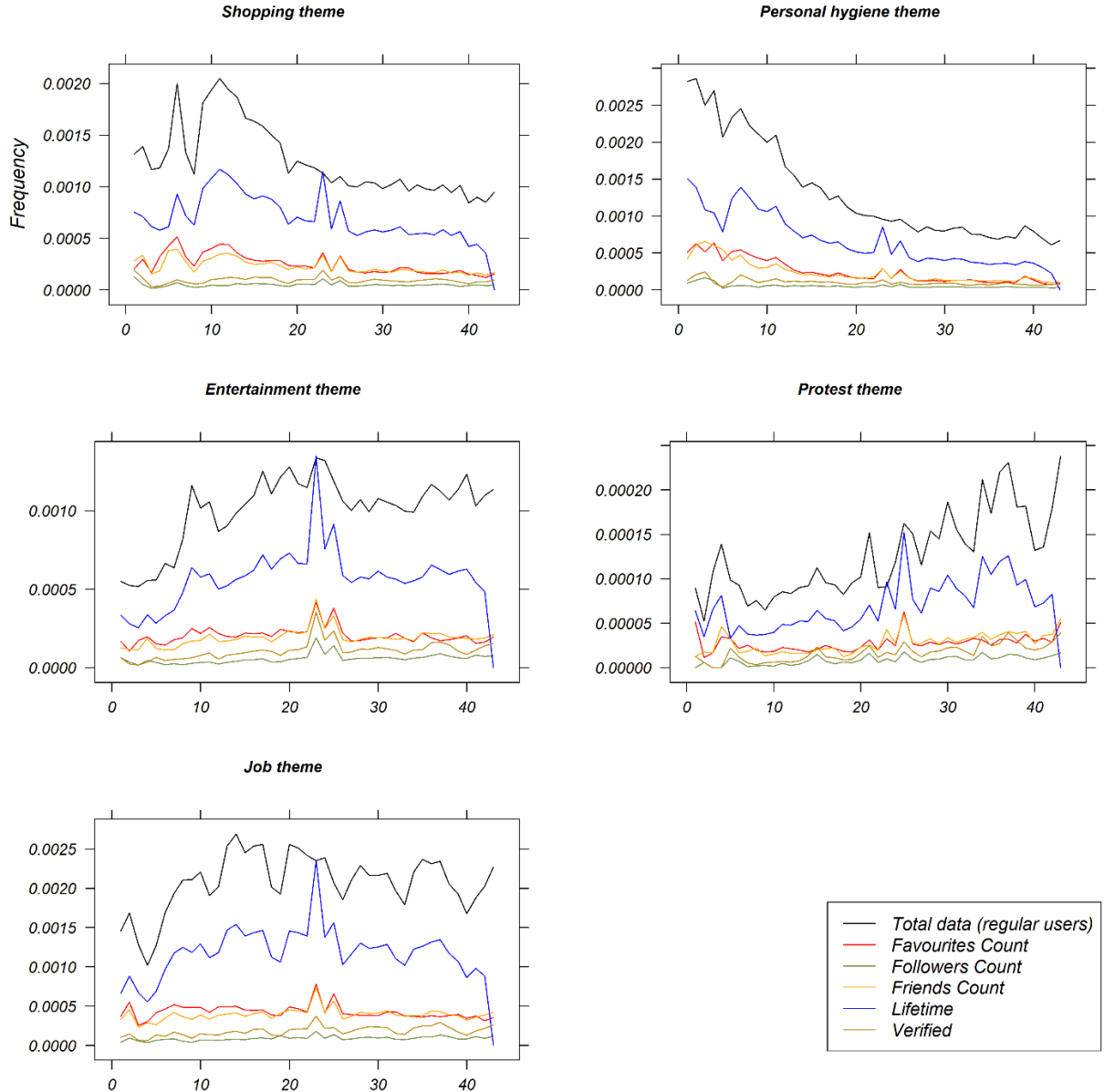

Fig 3. Word frequency of five main themes in the COVID-19 pandemic from 04-March-2020 to 15-April-2020

## 5 DISCUSSION OF THE RESULTS

Our analysis of a large dataset of user tweet texts reveals that user sentiments (emotions) in a critical situation (i.e., the COVID-19 pandemic) do not follow the network structure. Essentially, the results show that people do not tend to follow the network structure and the collective behavior of the network when it comes to vital issues such as jobs and protests. According to the network structure presented in Section 3 and Fig. 1, there is no relationship between user sentiment and network structure. Moreover, Fig. 3, shows that people followed an ascending trend in using protest theme words, while other user types do not have a significant increase in their direction. In contrast to people's sentiment and emotion, user behavior to some extent follows that of verified, experienced, and high degree centrality users. On the other hand, in matters of lower importance such as shopping, entertainment, and some other general issues such as personal hygiene, the results of the analysis show that users tended to follow collective behaviors. Therefore, this study suggests a new probability formula for user behavior in the network structure in Equation 4.

$$UB = f(li, v, fo, fr, fa)_t \tag{4}$$

Equation 4 shows that user behavior in the COVID-19 era for five themes can be modeled according to one or a combination of user characteristics. However, it should be noted that the weight of all user characteristics is not the same, and the experimental results demonstrate that people mostly follow verified users' behaviors, while the pattern of their behavior changes over time. To show the similarity of each theme with five users' type with respect to time, we used Z-Normalization y-offset shifting technique which is used in time series similarity analysis (Gogolou et al., 2018), in fact we tried to find how user behaviour (which depicts in words frequency in user's Tweet text) trend regarding the above-mentioned five themes matches (is similar to) the famous user types.

Equation 5 is used to compute the Z-Normalization y-offset:

$$Z = \frac{x_i - \mu}{\sigma} \tag{5}$$

Where Z is standard value, $x_i$ is observed value in each time point, $\mu$ is mean and $\sigma$ is standard deviation.

Figure 4 shows the similarity (matching) between the frequency of words in regular users' Tweet text and (a) the frequency of words in high favorite users' Tweet text (favorites count), (b) followers count (most followed users), (c) friends count (high degree centrality users ), (d) lifespan (experienced users), and (e) verified users based on Z-Normalization y-offset shifting value of word frequency of five main themes in the COVID-19 pandemic from 04-March-2020 to 15-April-2020. This figure highlights the role of time in Eq 4, for example, Fig 4 (a) shows the similarity of the trend between the frequency of the words of regular users' tweet text and users who have most favorite tweets (active users), as can be seen, the personal hygiene theme most of the time in the desired range is matched with this characteristic. However, Fig 4 shows that the relationship between user behavior in different themes and network characteristics may be different. For example, in Fig 4 (b), there is not much similarity between user behavior regarding health issues and users who have many followers that is why in Eq. 3 we found user behavior as a function of user characteristics and time, meaning that in some cases Eq. 3 can be modeled based on specific characteristics or a combination of them. Studying the similarities between different themes and users' characteristics reveals how people were affected by society during this pandemic. For example, the relationship between the entertainment theme and the number of friends in Fig 4 (c), or the very similarity between the protest theme and the verified users in Fig 4 (e). It seems that analyzing the similarity of users' behavior and reactions of users and network characteristics in times of crisis can lead to accurate predictions of the future behavior of people in society.

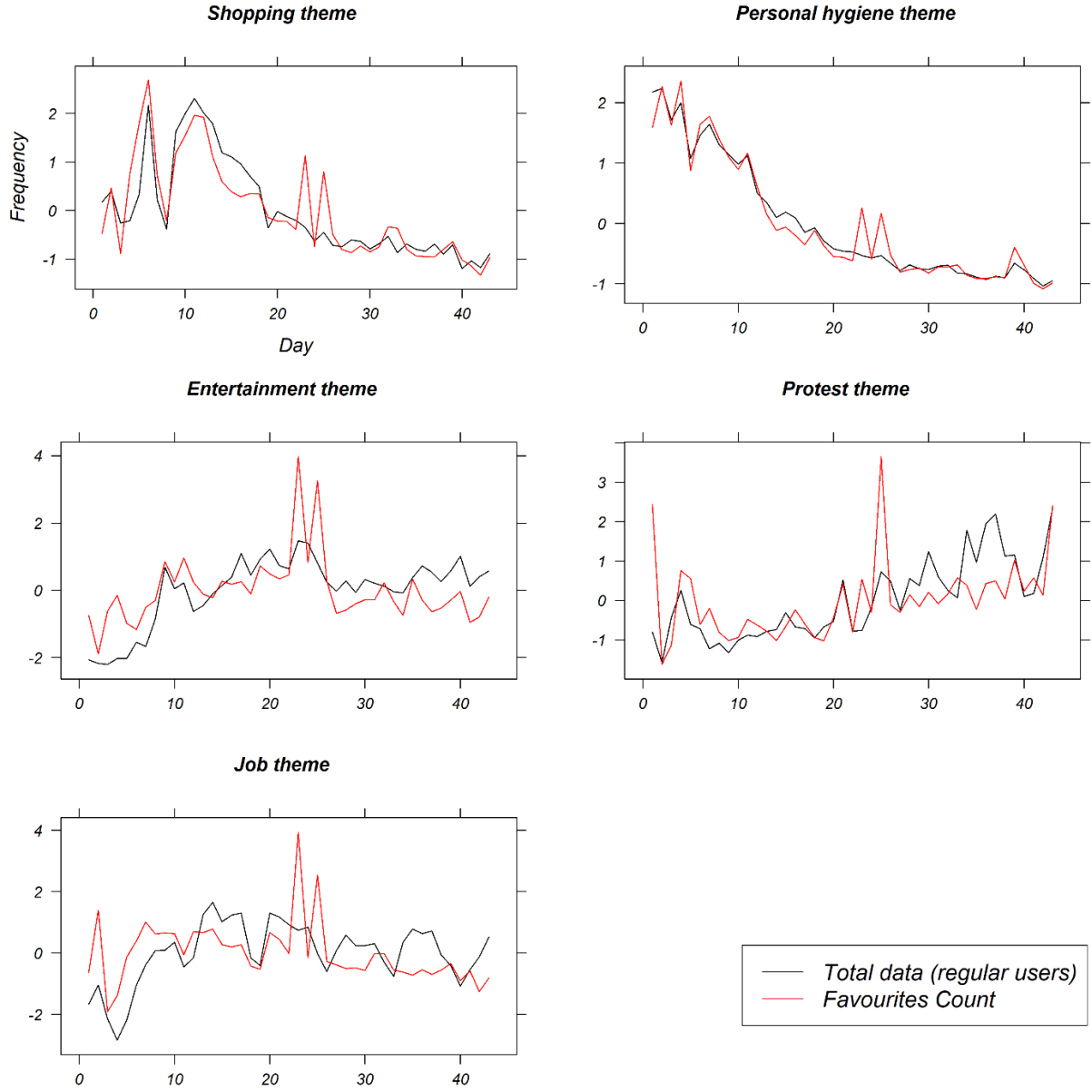

(a)

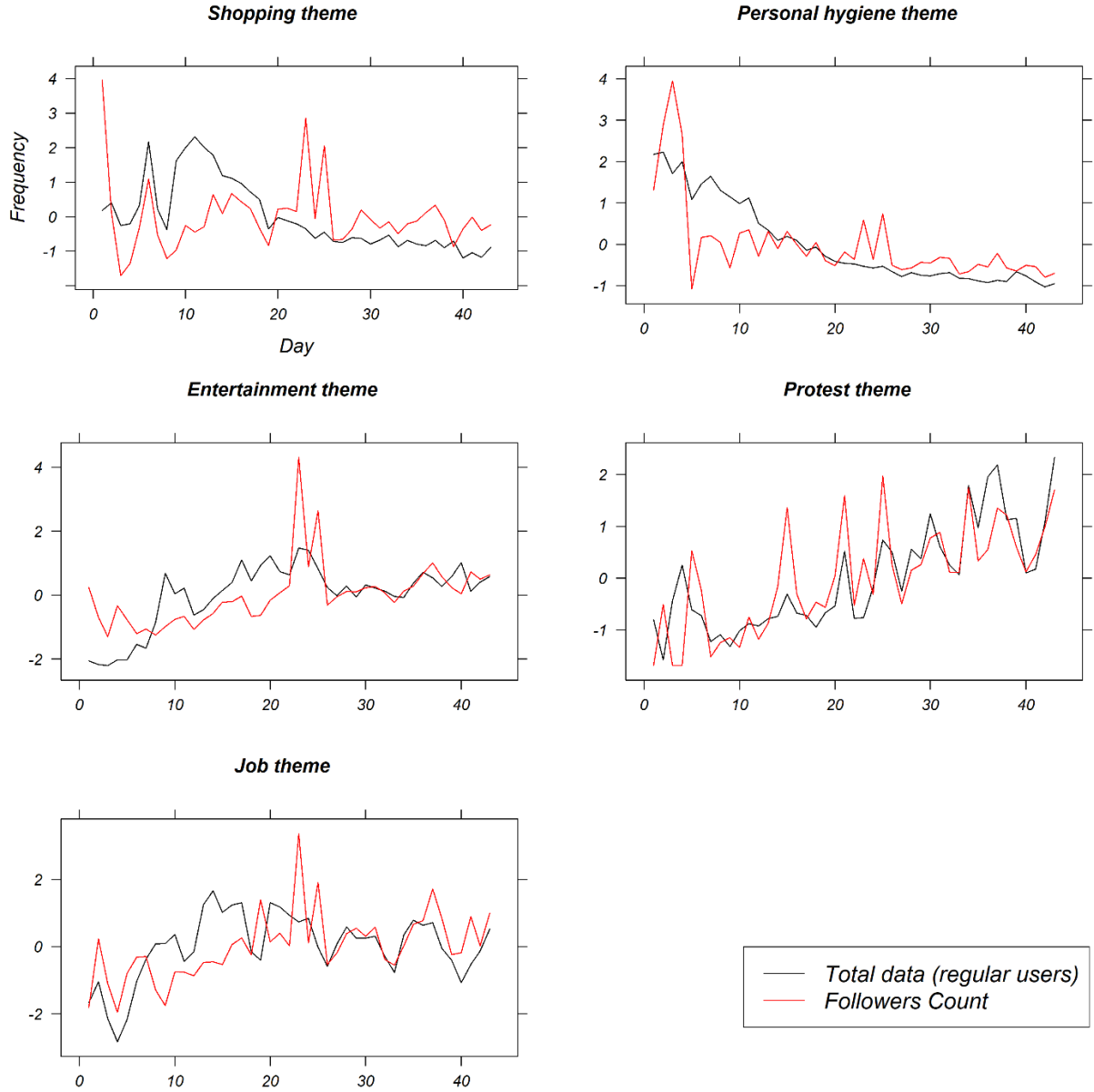

(b)

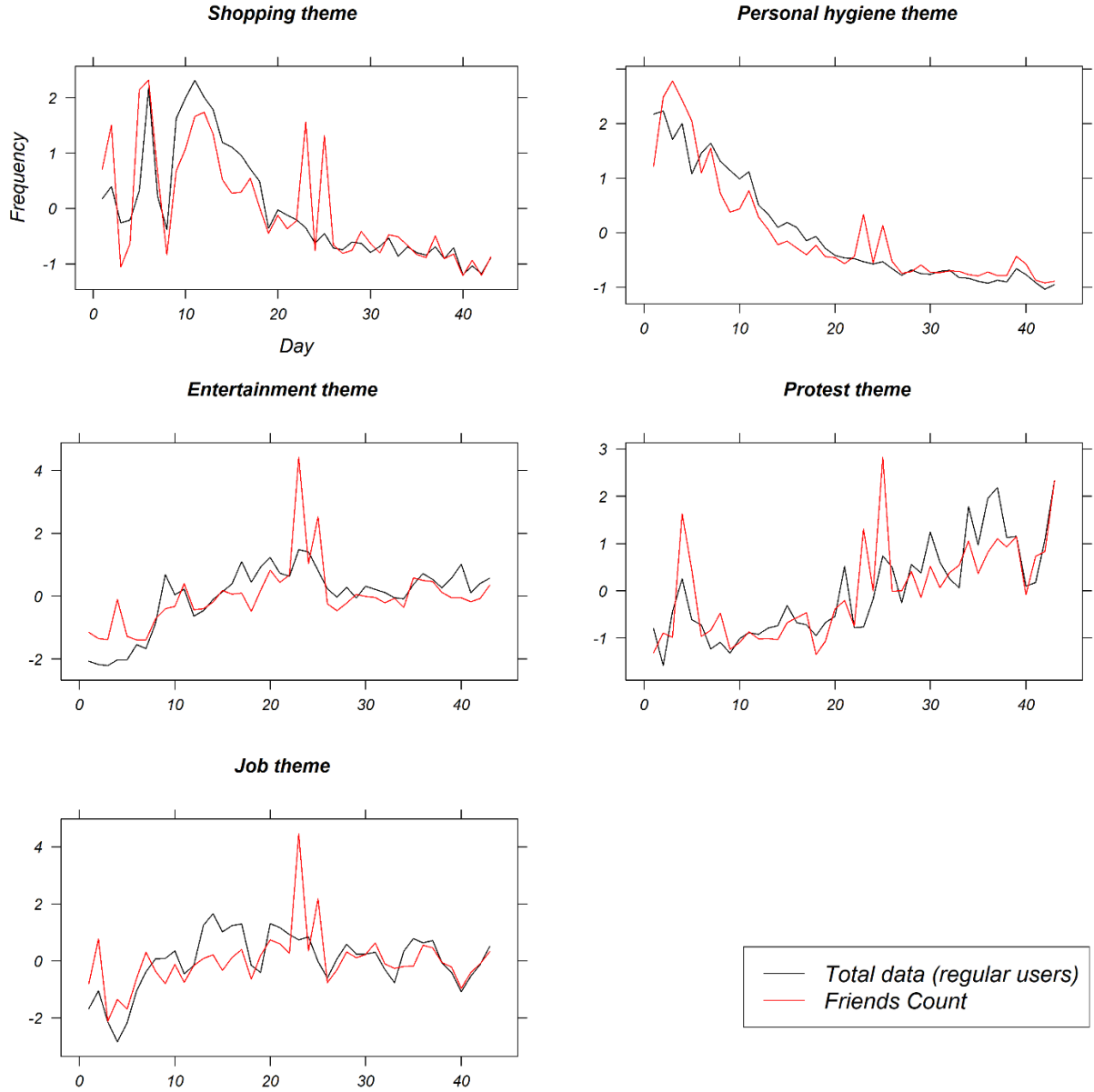

(c)

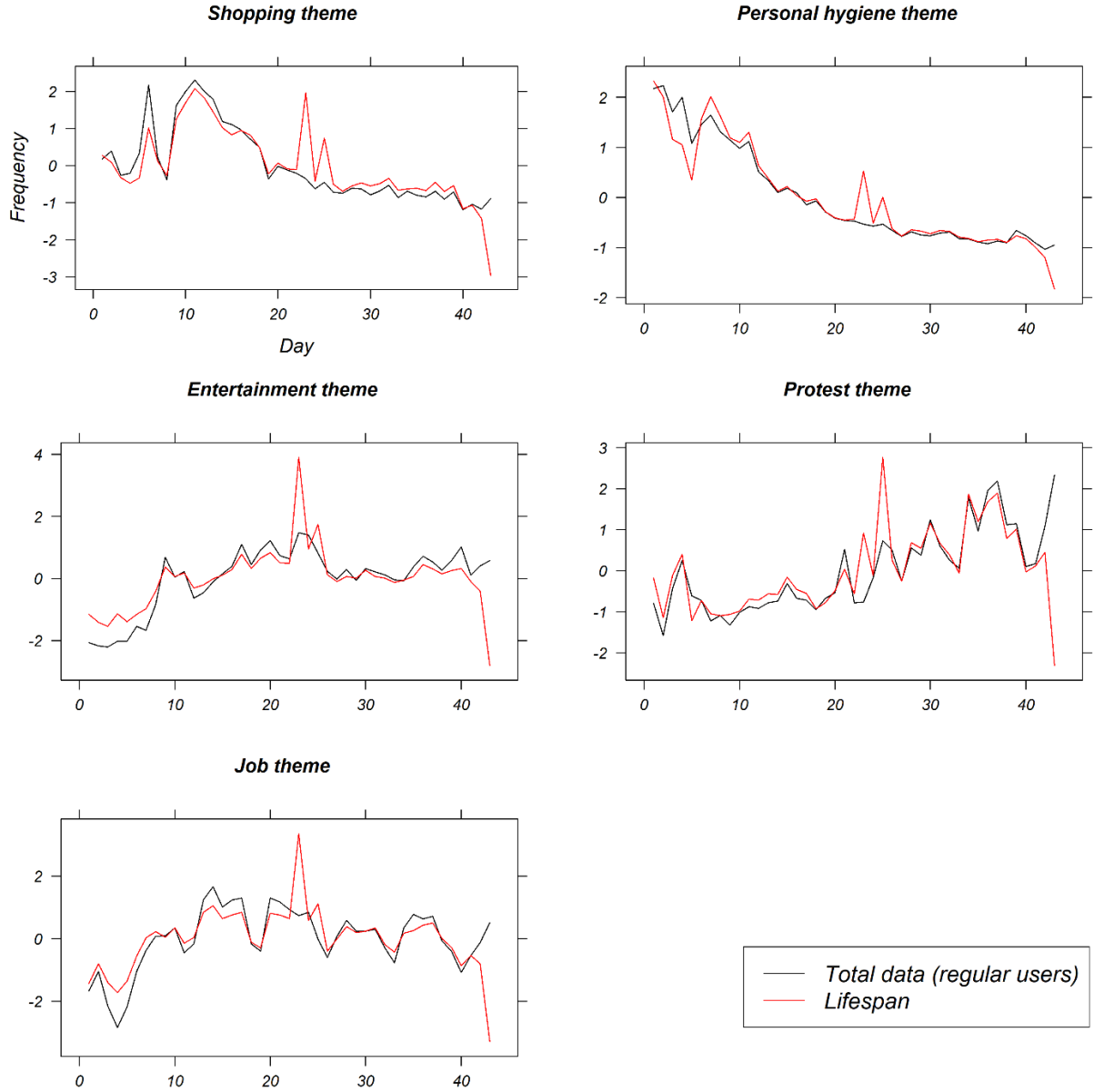

(d)

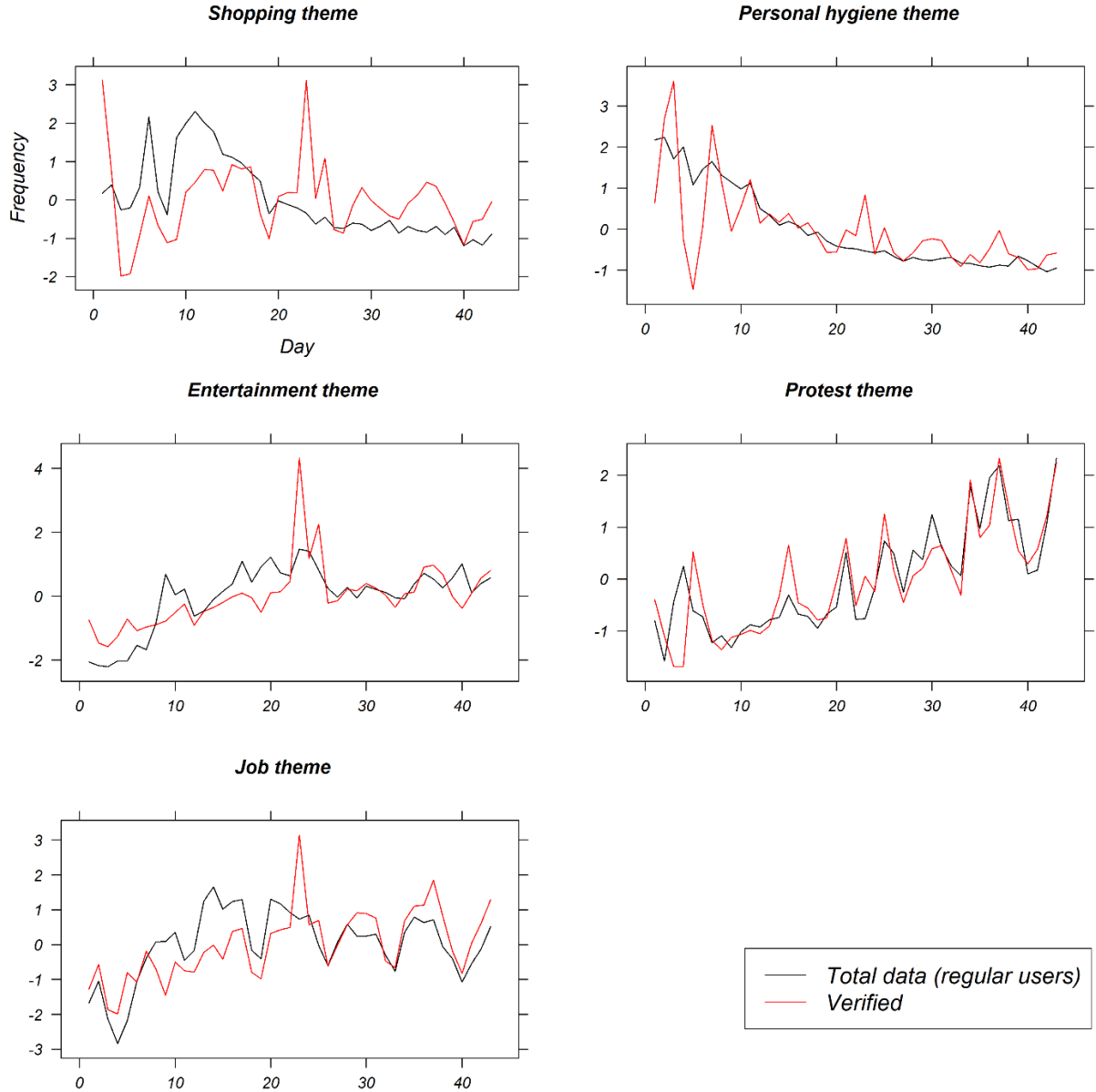

(e)

Fig 4. The similarity (matching) between the frequency of the words of regular users and (a) favorites count, (b) flowers count (most followed users), (c) friends count (high degree centrality users ), (d) lifespan (experienced users), and (e) verified users based on Z-Normalization y-offset shifting value of word frequency of five main themes in the COVID-19 pandemic from 04-March-2020 to 15-April-2020

This study uses the regression function to model the frequency of words of users' Tweet text regarding a specific theme based on the frequency of words of a particular user's type Tweet text. In this way, user behavior is a dependent variable, and users' type (user characteristics) is an independent variable. The results reveal that user behavior can model by each particular users' type. For instance, Fig. 5 shows that the regression variables of the protest theme and the verified users. As can be seen, the behavior of users in protesting fits well with the behavior of verified users by around 85% of multiple R-value and an insignificant standard error around 0.5.

SUMMARY OUTPUT

*Regression Statistics*

| | |
|---|---|
| Multiple R | 0.853068614 |
| R Square | 0.72772606 |
| Adjusted R Square | 0.721085233 |
| Standard Error | 0.528123808 |
| Observations | 43 |

ANOVA

| | df | SS | MS | F | Significance F |
|---|---|---|---|---|---|
| Regression | 1 | 30.56449 | 30.56449 | 109.5836 | 3.76562E-13 |
| Residual | 41 | 11.43551 | 0.278915 | | |
| Total | 42 | 42 | | | |

| | Coefficients | Standard Error | t Stat | P-value | Lower 95% | Upper 95% | Lower 95.0% | Upper 95.0% |
|---|---|---|---|---|---|---|---|---|
| Intercept | 0.000000012 | 0.080538 | 1.54E-07 | 1 | -0.162650032 | 0.16265006 | -0.16265003 | 0.162650057 |
| verified | 0.853068587 | 0.081491 | 10.46822 | 3.77E-13 | 0.688493625 | 1.01764355 | 0.68849363 | 1.01764355 |

figure 5. Regression of frequency of users' Tweet words regarding protest theme and the frequency of verified users Tweet words

Equation 6 is the relationship between user behavior regarding protest theme and verified users' behavior; this relationship is leaner, which shows the amount of similarity, and high level of dependency between these two variables.

$$UB_{protest} = 0.85 * v \tag{6}$$

Figure 6 indicates how the personal hygiene theme is fitted based on high degree centrality users (number of friends) by 91% of multiple R-value and an insignificant standard error around 0.4.

SUMMARY OUTPUT

*Regression Statistics*

| | |
|---|---|
| Multiple R | 0.919083089 |
| R Square | 0.844713724 |
| Adjusted R Square | 0.840926254 |
| Standard Error | 0.398840477 |
| Observations | 43 |

ANOVA

| | df | SS | MS | F | Significance F |
|---|---|---|---|---|---|
| Regression | 1 | 35.47797 | 35.47797 | 223.028484 | 3.51508E-18 |
| Residual | 41 | 6.522023 | 0.159074 | | |
| Total | 42 | 41.99999 | | | |

| | Coefficients | Standard Error | t Stat | P-value | Lower 95% | Upper 95% | Lower 95.0% | Upper 95.0% |
|---|---|---|---|---|---|---|---|---|
| Intercept | -0.00000004 | 0.060823 | -6E-07 | 0.99999953 | -0.12283377 | 0.122833702 | -0.12283377 | 0.122833702 |
| fr | 0.919083067 | 0.061542 | 14.93414 | 3.5151E-18 | 0.794795621 | 1.043370514 | 0.794795621 | 1.043370514 |

Figure 6. Regression of frequency of users' Tweet words regarding personal hygiene theme and the frequency of Tweet words of high degree centrality users (friends count)

Equation 7 is the relationship between user behavior regarding the personal hygiene theme and behavior of high degree centrality (fr) users; this relation also is leaner.

$$UB_{\text{Personal hygiene}} = 0.9 * fr \tag{7}$$

However, if we can find the time intervals in which the network behavior changes (Mahmoudi et al., 2018), then we can get a more accurate estimate and a more fitted regression function. In this vein, Fig. 7 shows the regression variables in a time interval (first 21 days) of the frequency of words regarding shopping theme. There is a more fitted regression between this theme and the frequency of words of experienced users (lifespan) regarding mentioned theme by around 95% multiple R-value and an insignificant standard error around 0.3.

SUMMARY OUTPUT

*Regression Statistics*

| | |
|---|---|
| Multiple R | 0.947127543 |
| R Square | 0.897050582 |
| Adjusted R Square | 0.891632192 |
| Standard Error | 0.29986762 |
| Observations | 21 |

ANOVA

| | df | SS | MS | F | Significance F |
|---|---|---|---|---|---|
| Regression | 1 | 14.88695179 | 14.88695 | 165.5567 | 7.90724E-11 |
| Residual | 19 | 1.708491198 | 0.089921 | | |
| Total | 20 | 16.59544299 | | | |

| | Coefficients | Standard Error | t Stat | P-value | Lower 95% | Upper 95% | Lower 95.0% | Upper 95.0% |
|---|---|---|---|---|---|---|---|---|
| Intercept | 0.124715971 | 0.08235527 | 1.514365 | 0.146392 | -0.047655589 | 0.297087532 | -0.04765559 | 0.29708753 |
| li | 1.099755405 | 0.085471787 | 12.86688 | 7.91E-11 | 0.920860898 | 1.278649912 | 0.9208609 | 1.27864991 |

Figure 7. Regression of frequency of users' Tweet words regarding shopping theme and the frequency of Tweet words of experienced users (li) in a time interval (first 21 days)

Equation 8 shows that the suggested regression function for Fig 7. This relation also is leaner, which indicates a very close relationship between two variables.

$$UB_{shopping} = 1.1 * li + 0.12 \tag{8}$$

Another interesting user behavior identified in this study is that users are more likely to follow other people's behavior when they feel negative. Figures 1 and 3 show that in the initial days of the pandemic when the users' sentiments are mostly negative, their behavior is very close to other targeted groups of users. Indeed, in Fig. 3, the plot of the five aforementioned themes associated with the plot of the five main user characteristics shows that user behaviors are very similar to each other until March 13th. This phenomenon demonstrates that people's behavior (not sentiment) in critical situations tend to follow influential users; however, it is the role of the psychologist to identify the causation(s) of this phenomenon.

## 6   CONCLUSION AND FUTURE DIRECTIONS

This study analyzed around 18 million tweet texts that were posted during the early period of the COVID-19 pandemic. First, the user sentiments based on time were extracted, and then, according to five specified themes (shopping, personal hygiene, entertainment, protests, and job), user behaviors during the COVID-19 era were identified. Based on the results of the analysis, this study proposed a probability function based on user characteristics, such as the number of followers, degree of centrality, verified user status, favorites count, and using experience. The results showed that (1) users sentiments do not follow those of influential users during the critical situation, (2) users' sentiments are a function of time, and (3) in contrast to the users sentiments, some user behaviors follow those of the influential and verified users, especially when the mood of society is generally negative. In addition, this study showed that the sentiment analysis of textual data can accurately depict social phenomena and show the trajectory of people's behavior in times of crisis, such as the COVID-19 pandemic. Moreover, this research confirmed the relevancy of social physics theory and the role of this theory in formulating social phenomena using big data.

This research on people's behaviors at this challenging time has been a useful experience and provides insights for the future that can hopefully be used to solve crises quickly, get ahead of the curve, and make better decisions. Further assessment and testing of the proposed formula for predicting user behaviors in a crisis is a promising research avenue. In addition, analyzing the effect of geo-location and movement on the COVID-19 pandemic is still an area of great interest and debate, not least because it is essential to investigate the different behaviors of different communities during this pandemic in order to develop effective policies and strategies.

**REFERENCES**

Barkur, G., Vibha, & Kamath, G. B. (2020). Sentiment analysis of nationwide lockdown due to COVID 19 outbreak: Evidence from India. *Asian Journal of Psychiatry*, *51*, 102089. https://doi.org/10.1016/j.ajp.2020.102089

Bhattacharya, K., & Kaski, K. (2018). *arXiv : 1804 . 04907v1 [ physics . soc-ph ] 13 Apr 2018 Social Physics : Uncovering Human Behaviour from Communication*. April. https://doi.org/10.1080/23746149.2018.1527723

Carothers, T., & Press, B. (2020). *The Global Rise of Anti-Lockdown Protests—and What to Do About It*. https://www.worldpoliticsreview.com/articles/29137/amid-the-covid-19-pandemic-protest-movements-challenge-lockdowns-worldwide

Chuang, Y. C., Huang, Y. L., Tseng, K. C., Yen, C. H., & Yang, L. H. (2015). Social capital and health-protective behavior intentions in an influenza pandemic. *PLoS ONE*, *10*(4), 1–14. https://doi.org/10.1371/journal.pone.0122970

Cinelli, M., Quattrociocchi, W., Galeazzi, A., Valensise, C. M., Brugnoli, E., Schmidt, A. L., Zola, P., Zollo, F., & Scala, A. (2020). *The COVID-19 Social Media Infodemic*. 1–18. http://arxiv.org/abs/2003.05004

da Silva, N. F. F., Coletta, L. F. S., Hruschka, E. R., & Hruschka Jr., E. R. (2016). Using unsupervised information to improve semi-supervised tweet sentiment classification. *Information Sciences*, *355–356*, 348–365. https://doi.org/10.1016/j.ins.2016.02.002

David Nettleton. (2014). Analysis of Data on the Internet III – Online Social Network Analysis. In *Commercial Data Mining* (pp. e27–e42). https://doi.org/https://doi.org/10.1016/B978-0-12-416602-8.00016-9

Depoux, A., Martin, S., Karafillakis, E., Bsd, R. P., Wilder-Smith, A., & Larson, H. (2020). The pandemic of social media panic travels faster than the COVID-19 outbreak. *Journal of Travel Medicine*. https://doi.org/10.1093/jtm/taaa031

Dictionary, C. (2020). *Cambridge*. https://dictionary.cambridge.org/dictionary/english/sentiment

Dubey, A. D. (2020). Twitter Sentiment Analysis during COVID19 Outbreak. *SSRN Electronic Journal*, *March*, 1–9. https://doi.org/10.2139/ssrn.3572023

Fegert, J. M., Vitiello, B., Plener, P. L., & Clemens, V. (2020). Challenges and burden of the Coronavirus 2019 (COVID-19) pandemic for child and adolescent mental health: A narrative review to highlight clinical and research needs in the acute phase and the long return to normality. *Child and Adolescent Psychiatry and Mental Health*, *14*(1), 1–11. https://doi.org/10.1186/s13034-020-00329-3

Feldman, & Ronen. (2013). Techniques and applications for sentiment analysis. *Communications of the ACM*, *56*(4).

Ferlander, S. (2007). The Importance of Different Forms of Social Capital for Health. *Acta Sociologica*, *50*(2), 115–128. https://doi.org/10.1177/0001699307077654

Fornacciari, P., Mordonini, M., & Tomauiolo, M. (2015). Social network and sentiment analysis on twitter: Towards a combined approach. *CEUR Workshop Proceedings*, *1489*, 53–64.

Freeman, L. C. (1978). Centrality in social networks conceptual clarification. *Social Networks*, *1*(3), 215–239. https://doi.org/10.1016/0378-8733(78)90021-7

Giachanou, A., & Crestani, F. (2016). Like it or not: A survey of Twitter sentiment analysis methods. *ACM Computing Surveys*, *49*(2). https://doi.org/10.1145/2938640

Gogolou, A., Tsandilas, T., Palpanas, T., Bezerianos, A., Similarity, C., Gogolou, A., Tsandilas, T., Palpanas, T., & Bezerianos, A. (2018). Visualizations To cite this version : HAL Id : hal-01845008 Comparing Similarity Perception in Time Series Visualizations. *IEEE Transactions on Visualization and Computer Graphics*.

Gong, L., & Wang, H. (2018). *When Sentiment Analysis Meets Social Network*. 1455–1464. https://doi.org/10.1145/3219819.3220120

Gui, L., Zhou, Y., Xu, R., He, Y., & Lu, Q. (2017). Learning representations from heterogeneous network for sentiment classification of product reviews. *Knowledge-Based Systems*, *124*, 34–45. https://doi.org/10.1016/j.knosys.2017.02.030

Jin, S., & Zafarani, R. (2018). Sentiment Prediction in Social Networks. *2018 IEEE International Conference on Data Mining Workshops (ICDMW)*, 1340–1347. https://doi.org/10.1109/ICDMW.2018.00190

Kaggle. (2020). *Coronavirus (covid19) Tweets*. Kaggle. https://www.kaggle.com/smid80/coronavirus-covid19-tweets-early-april

Kim, B. (2020). Effects of Social Grooming on Incivility in COVID-19. *Cyberpsychology, Behavior and Social Networking*, *X*(X), 1–7. https://doi.org/10.1089/cyber.2020.0201

Limaye, R. J., Sauer, M., Ali, J., Bernstein, J., Wahl, B., Barnhill, A., & Labrique, A. (2020). Building trust while influencing online COVID-19 content in the social media world. *The Lancet Digital Health*, *2019*(20), 2019–2020. https://doi.org/10.1016/S2589-7500(20)30084-4

Mahmoudi, A., Bakar, A. A., Sookhak, M., & Yaakub, M. R. (2020). A Temporal User Attribute-Based Algorithm to Detect Communities in Online Social Networks. *IEEE Access*, *8*, 154363–154381. https://doi.org/10.1109/ACCESS.2020.3018941

Mahmoudi, A., Yaakub, M. R., & Abu Bakar, A. (2018). A new method to discretize time to identify the milestones of online social networks. *Social Network Analysis and Mining*, *8*(1), 1–20. https://doi.org/10.1007/s13278-018-0511-4

Pêcher, C., Lemercier, C., & Cellier, J. M. (2011). The influence of emotions on driving behavior. *Traffic Psychology: An International Perspective*, *June 2015*, 145–158.

Pentland, A. (2015). *It is capable of revealing the structure and behavioural patterns of social systems at different scales from individual to societal level as well as capturing the long term evolution of the society*.

Pitas, N., & Ehmer, C. (2020). Social Capital in the Response to COVID-19. *American Journal of Health Promotion : AJHP*, 890117120924531. https://doi.org/10.1177/0890117120924531

Rinker, T. (2019). *Package "sentimentr": Calculate Text Polarity Sentiment*. 59. https://cran.r-project.org/web/packages/sentimentr/sentimentr.pdf

Rosenberg, H., Syed, S., & Rezaie, S. (2020). The twitter pandemic: The critical role of twitter in the dissemination of medical information and misinformation during the COVID-19 Pandemic. *Cjem*, *May*, 1–7. https://doi.org/10.1017/cem.2020.361

Schober, P., Boer, C., & Schwarte, L. A. (2018). *Correlation Coefficients: Appropriate Use and Interpretation*. *126*(5), 1763–

1768. https://doi.org/10.1213/ANE.0000000000002864

Taboada, M., Brooke, J., & Voll, K. (2011). Lexicon-Based Methods for Sentiment Analysis. *American Psychological Society*, *August 2010*. http://www.jstor.org/stable/20182753%0D

Tan, C., Lee, L., & Tang, J. (2013). *User-Level Sentiment Analysis Incorporating Social Networks Background : Sentiment Analysis ( SA )*. *March*, 1397–1405.

Vishal.A.Kharde, & Sonawane, S. S. (2016). Sentiment Analysis of Twitter Data: A Survey of Techniques. *International Journal of Computer Applications*, *139*(11), 5–15. https://doi.org/10.5120/ijca2016908625

West, R., Paskov, H. S., Leskovec, J., & Potts, C. (2014). Exploiting Social Network Structure for Person-to-Person Sentiment Analysis. *Transactions of the Association for Computational Linguistics*, *2*, 297–310. https://doi.org/10.1162/tacl_a_00184

WHO. (2020). *WHO*. https://www.who.int/emergencies/diseases/novel-coronavirus-2019/advice-for-public

Wiederhold, B. K. (2020). Social Media Use During Social Distancing. *Cyberpsychology, Behavior and Social Networking*, *23*(5), 275–277. https://doi.org/10.1089/cyber.2020.29181.bkw

Yang, Y., Chang, M. W., & Eisenstein, J. (2016). Toward socially-infused information extraction: Embedding authors, mentions, and entities. *EMNLP 2016 - Conference on Empirical Methods in Natural Language Processing, Proceedings*, 1452–1461. https://doi.org/10.18653/v1/d16-1152

Yang, Y., & Eisenstein, J. (2017). Overcoming Language Variation in Sentiment Analysis with Social Attention. *Transactions of the Association for Computational Linguistics*, *5*, 295–307. https://doi.org/10.1162/tacl_a_00062

Yoo, S. Y., Song, J. I., & Jeong, O. R. (2018). Social media contents based sentiment analysis and prediction system. *Expert Systems with Applications*, *105*, 102–111. https://doi.org/10.1016/j.eswa.2018.03.055

Zou, X., Yang, J., & Zhang, J. (2018). Microblog sentiment analysis using social and topic context. *PLoS ONE*, *13*(2), 1–24. https://doi.org/10.1371/journal.pone.0191163